\documentstyle[aps,preprint,tighten]{revtex}

\def\be{\begin{equation}}
\def\ee{\end{equation}}
\def\bea{\begin{eqnarray}}
\def\eea{\end{eqnarray}}
\newcommand{\lsim}{\mathrel{\mathop{\kern 0pt \rlap
  {\raise.2ex\hbox{$<$}}}
  \lower.9ex\hbox{\kern-.190em $\sim$}}}
\newcommand{\gsim}{\mathrel{\mathop{\kern 0pt \rlap
  {\raise.2ex\hbox{$>$}}}
  \lower.9ex\hbox{\kern-.190em $\sim$}}}
\newcommand{\gagamma}{g_{a\gamma\gamma}}

\begin{document}

\title{Prospects of solar axion searches
        with crystal detectors}

\author{ S. Cebri\'{a}n, E. Garc\'{\i}a,
         D. Gonz\'{a}lez, I. G. Irastorza, A. Morales, J. Morales,\\ A. Ortiz de
         Sol\'{o}rzano,
         J. Puimed\'{o}n, A. Salinas, M. L. Sarsa, S. Scopel, J. A. Villar}

\address{Laboratorio de F\'{\i}sica Nuclear. Universidad de Zaragoza\\
              50009, Zaragoza, SPAIN}

\maketitle\abstract{
A calculation of the expected signal due to Primakoff coherent
conversion of solar axions into photons via Bragg scattering in
several crystal detectors is presented. The results are confronted
with the experimental sensitivities of present and future
experiments concluding that the sensitivity of crystal
detectors does not challenge the globular cluster limit on the axion--photon
coupling $\gagamma$.
In particular, in the
axion mass window $m_a\gsim 0.03$ eV explored with this technique
(not accessible at present by
other methods)
$\gagamma$ might be
constrained down to 10$^{-9}$ GeV$^{-1}$ (the recent
helioseismological bound) provided that significant improvements
in the parameters and performances of these detectors be achieved
and
large statistics accumulated. This bound should be considered as a
minimal goal for the sensitivity of future crystal experiments.
Consequently, finding a positive signal at this level of sensitivity would
necessarily imply revisiting other more stringent astrophysical
limits derived for the same range of $m_a$ values.
}
\vspace{2 cm}


\newpage

\section{Introduction}
Introduced twenty years ago as the Nambu--Goldstone boson of the Peccei--Quinn symmetry
to explain in an elegant way CP conservation in QCD \cite{PQ}, the axion
is remarkably also one of the best candidates
to provide at least a fraction of the Dark Matter needed in Cosmology
in order to explain both gravitational measurements and models of structure formation.

Axion phenomenology is determined by its mass $m_a$
which in turn is fixed by the scale $f_a$ of the Peccei--Quinn symmetry
breaking, $m_a$ $\simeq$ 0.62 eV $(10^7$ ${\rm GeV}/f_a)$.
No hint is provided by theory about where the $f_a$ scale should
be.
A combination of astrophysical and nuclear physics
constraints, and the requirement that the axion relic abundance
does not overclose the Universe, restricts the allowed range of
viable axion masses into a relatively narrow window\cite{kolbturner,sn87a,cerenkov}:
\begin{eqnarray}
 10^{-6}  {\rm eV}  \lsim & m_a & \lsim 10^{-2} {\rm eV} \nonumber \\
3  \;{\rm eV} \lsim  & m_a &  \lsim  20\;  {\rm eV}.
\label{eq:limits}
\end{eqnarray}
The physical process used in axion search experiments
is the Primakoff effect. It makes use of the coupling between the
axion field $\psi_a$ and the electromagnetic tensor:
\begin{equation}
{\cal L}=-\frac{1}{4}g_{a\gamma \gamma}\psi_a\epsilon_{\mu \nu \alpha
\beta}F^{\mu \nu} F^{\alpha \beta}=-
\gagamma \psi_a \vec{B}\cdot\vec{E}
\label{eq:lagrangiana}
\end{equation}
\noindent
and allows for the conversion of the axion into a photon.

Like all the other axion couplings, $\gagamma$ is proportional to
$m_a$\cite{kolbturner,raffelt_review}:
\begin{equation}
\gagamma\simeq 0.19 \frac{m_a}{\rm eV} \left(\frac{E}{N} -
\frac{2(4+z)}{3(1+z)} \right) 10^{-9}{\rm GeV}^{-1}
\label{eq:gagamma}
\end{equation}
\noindent where $E/N$ is the PQ symmetry anomaly
and the second term in parenthesis is the chiral symmetry breaking
correction.
The anomaly E/N depends on the particular axion model,
while the symmetry breaking correction is a
function of
the parameter $z\equiv m_u/m_d\simeq$ 0.56.
Two popular models are the GUT--DFSZ axion\cite{dfsz} (E/N=8/3) and the KSVZ
axion\cite{ksvz} (E/N=0).
However, it is possible to build viable axion models with different values
of E/N\cite{kim} and the determination of the parameter $z$ is
subject to some theoretical uncertainties\cite{moroi}.
This implies that a very small or even vanishing
$\gagamma$ cannot be in principle excluded.

Solid state detectors provide a simple mechanism for axion detection
\cite{zioutas,creswick}.
Axions can pass in the
proximity of the atomic nuclei of the crystal where the intense electric field can
trigger their conversion into photons.
In the process the energy of the outgoing photon is equal to that of the
incoming axion.

Axions can be efficiently produced in the interior of the Sun
by Primakoff conversion of the blackbody photons in the fluctuating electric field
of the plasma.
The resulting flux has an outgoing average axion energy
$E_a$ of about 4 keV (corresponding to
the temperature in the core of the Sun, $T\sim 10^7 K$)
that can produce detectable x--rays in a crystal detector.
Depending on the direction of the incoming axion flux with
respect to the planes of the crystal lattice, a coherent effect can be produced
when the Bragg condition is fulfilled,
leading so to a strong enhancement of the
signal. A correlation of the expected count--rate
with the position of the Sun in the sky is a distinctive signature of the axion
which can be used, at the least, to improve the signal/background
ratio.

The process described above is independent on
$m_a$ and so are the achievable bounds
for the axion--photon coupling
$\gagamma$.
This fact is particularly appealing, since
other experimental techniques
are limited to a more restricted mass range:
``haloscopes''\cite{cavities1}, that use electromagnetic cavities
to look for the resonant conversion
into microwaves of non relativistic cosmological dark halo axions,
do not extend their search beyond $m_a\simeq 50\;\mu{\rm eV}$, while
the dipole magnets used in ``helioscope''\cite{tokio1} experiments are
not sensitive at present to solar axions heavier than $m_a\simeq 0.03\;{\rm
eV}$.

A pilot experiment carried out by the SOLAX Collaboration\cite{cosmesur}
has already searched for axion
Primakoff conversion in a germanium crystal of 1 kg obtaining the
limit $\gagamma\lsim 2.7 \times 10^{-9}\; {\rm GeV}^{-1}$.
This is the (mass independent but solar model dependent) most
stringent laboratory bound for the axion--photon coupling obtained
so far, although less restrictive than the globular cluster
bound\cite{hb} $\gagamma\lsim 0.6 \times 10^{-10}\; {\rm GeV}^{-1}$.
Notice however that the experimental
accuracy of solar observations is orders of
magnitude better than for any other star.

As a matter of fact, the solar model itself already requires\cite{helio}
$\gagamma\lsim 10^{-9}$ ${\rm GeV}^{-1}$, whereas the above mentioned Ge
experiment has not yet reached such sensitivity.
The $10^{-9} {\rm GeV^{-1}}$ limit
sets, then, a minimal goal for the sensitivity of future
experiments, prompting the need for a systematic discussion
of present efforts and future prospects for axion searches with crystals.
In the following we give the result of such an analysis,
focusing on Germanium, ${\rm TeO}_2$ and NaI detectors.

\section{Primakoff conversion in crystals}
\label{sec:theory}
We will make use of the calculation of the flux of solar axions
of Ref. \cite{raffelt},
with no direct coupling between the axion and
the leptons ({\it hadronic axion}),
with the modifications introduced in Ref.
\cite{creswick} to include helium and metal diffusion in the solar
model.
A useful parametrization of the flux is the following:
\begin{equation}
\frac{d\Phi}{dE_a}=\sqrt{\lambda}\frac{\Phi_0}{E_0}\frac{(E_a/E_0)^3}{e^{E_a/E_0}-1}
\label{eq:flux}
\end{equation}
\noindent where
$\lambda$=($\gagamma\times 10^8/$${\rm GeV}$$^{-1}$)$^4$
is a dimensionless coupling introduced for later convenience,
$\Phi_0$=5.95 $\times$$10^{14}$ cm$^{-2}$
sec$^{-1}$ and $E_0$=1.103 keV.

The expected count-rate in a solid-state ionization detector,
integrated in the electron-equivalent energy window
$E_1$$<$$E_{ee}$$<$$E_2$, is calculated starting from the expression:
\begin{equation}
R(E_1,E_2)=\int_{E_1}^{E_2}d E_{ee} \int_0^{\infty} d E_{\gamma}
\frac{d R}{d E_{\gamma}}(E_{\gamma})
\frac{1}{\Delta\sqrt{2\pi}}
e^{-\frac{(E_{ee}-E_{\gamma})^2}{2 \Delta^2}}
\label{eq:rate}
\end{equation}
\noindent where
$E_{\gamma}$ is the energy of the outcoming
photon, $\Delta$ is the resolution of the detector, FWHM $\simeq$ 2.35 $\Delta$, while:
\begin{equation}
\frac{d R}{d E_{\gamma} } (E_{\gamma}) =\int_0^{\infty} d E_a \frac{d \Phi}{d E_a}(E_a)
\int d\Omega \frac{d \sigma}{d \Omega d E_{\gamma}}(E_{\gamma}).
\label{eq:rate1}
\end{equation}
\noindent $\sigma$ is the cross section of the process.
The recoil of the nucleus can be neglected, so the energy of the
outgoing photon is equal to that of the incoming axion
and the differential cross section for the Primakoff conversion
may be written as:
\begin{equation}
\frac{d \sigma}{d E_{\gamma} d \Omega}=\frac{d \sigma}{d \Omega}
\delta(E_{\gamma}-E_a)
\end{equation}
\noindent where\cite{cross}:
\begin{equation}
\frac{d\sigma}{d\Omega}=\frac{\gagamma^2}{16 \pi^2}F_a^2(\vec{q})
\sin^2(2\theta).
\label{eq:cross}
\end{equation}
\noindent $2 \theta$ and is the scattering angle, while
\begin{equation}
F_a(\vec{q})=k^2 \int d^3x \;\phi(\vec{x})e^{i\vec{q}\cdot\vec{x}}
\label{eq:formfactor}
\end{equation}
\noindent is the Fourier transform of the electrostatic field
$\phi$. $\vec{q}$ is the transferred momentum,
$q\equiv|\vec{q}|=2 k \sin\theta$ and
$k$$\equiv$$|\vec{k}|$$\simeq$$E_a$ is the axion momentum.

The energy distribution of Eq.(\ref{eq:flux}) implies that
$q$ corresponds to a wavelength of a few ${\rm \AA}$, which is of the
order of the distances between atoms in a typical crystal lattice.
As a consequence of this a Bragg-reflection pattern arises in the calculation of
the Fourier transform of Eq.(\ref{eq:formfactor}) due to the
periodic properties of the electric field in the crystal.
Using translational invariance the periodic field may be written as\cite{cristalografia}
(we consider here the general case of a multi--target species):
\begin{equation}
\phi(\vec{x})=\sum_{ij} \phi_{ij}(\vec{x})=
\sum_{ij}\frac{Z_j e}{4\pi|\vec{x}-\vec{x}_i|}e^{-\frac{|\vec{x}-\vec{x}_i|}
{r_j}}=\sum_G
\;n_G e^{i\vec{G}\cdot\vec{x}}.
\label{eq:potential}
\end{equation}
\noindent where $\vec{x}_i$ are the positions in space of the elements
of the lattice, $Z_j$ is the atomic number of the $j$--th type target nucleus,
$r_j$
is the screening length of the corresponding atomic electric field parametrized
with a Yukawa--type potential,
and the sum in the last term is intended
over the vectors $\vec{G}$ of the reciprocal lattice, defined by the
property $\exp{i \vec{G}_i\cdot \vec{x}_i}\equiv 1$.
A multi--target crystal
is described by
a Bragg lattice with a basis whose sites are occupied
by atoms of different types. Indicating by $a_i^j$ the $i$'th basis vector
occupied by the
$j$'th target-nucleus type, the $n_G$ coefficients appearing in
Eq.(\ref{eq:potential}) turn out to be:

\begin{equation}
n_G=\frac{1}{v_a}\sum_j S_j(\vec{G}) F^0_{a,j}(\vec{q}=\vec{G})
\label{eq:ng}
\end{equation}
\noindent where
\begin{equation}
S_j(\vec{G})=\sum_i e^{i \vec{a}_i^j \vec{G}}
\label{eq:struct}
\end{equation}
\noindent are the structure functions of the crystal and
\begin{equation}
F^0_{a,j}(\vec{q})\equiv k^2\int d^3x\; \phi_{ij}(\vec{x}) e^{i\vec{q}\cdot\vec{x}}=
\frac{Z_j e k^2}{r_j^{-2}+q^2}.
\label{eq:f0}
\end{equation}
\noindent
is the form factor of Eq.(\ref{eq:formfactor}) for a single
j--type target.
$v_a$ indicates the volume of the elementary cell of the
lattice.

Integration of Eqs.(\ref{eq:rate},\ref{eq:rate1}) over $E_{ee}$ and $E_a$ leads to:
\begin{eqnarray}
&&R(E_1,E_2)=\int 2 c \frac{d^3 q}{q^2} \frac{d \Phi}{d
E_a}(E_a)\frac{\gagamma^2}{16 \pi^2} |F_a(\vec{q})|^2
\sin^2(2 \theta) \; {\cal W} =\label{eq:final_a}\\
&&=(2\pi)^3 2 c \hbar \frac{V}{v_a^2}\sum_G\frac{d \Phi}{d E_a}(E_a)
\frac{1}{|\vec{G}|^2}\frac{\gagamma^2}{16 \pi^2}
|\sum_j F_{a,j}^0(\vec{G})S_j(\vec{G})|^2  \sin^2(2 \theta)\;  {\cal W}
\label{eq:final_b}
\end{eqnarray}
\noindent where $V$ is the volume of the detector,
\begin{equation}
{\cal W}(E_1,E_2,E_a,\Delta)=\frac{1}{2}
\left[ {\rm erf}\left( \frac{E_a-E_1}{\sqrt{2}\Delta}\right)-
{\rm erf}\left(\frac{E_a-E_2}{\sqrt{2}\Delta} \right)
\right],
\end{equation}
\noindent
and in Eq.(\ref{eq:final_a}) $|F_a(\vec{q})|^2$ has been
expanded by making use of
Eqs.(\ref{eq:formfactor},\ref{eq:potential},\ref{eq:ng},\ref{eq:struct}).
In the final result the integral over the transferred momentum
has been replaced by a sum over the vectors of the reciprocal lattice,
i.e. over the peaks that are produced when the Primakoff conversion verifies the
Bragg condition $\vec{q}=\vec{G}$ and the crystal interacts in a coherent way.

The Bragg condition implies that in Eq.(\ref{eq:final_b})
$E_a=\hbar c \frac{|\vec{G}|^2}{2
\hat{u}\cdot \vec{G}}$ where the unitary vector $\hat{u}$ points toward
the Sun. This term induces a time dependence in the expected signal.

\section{Time correlation and background rejection}
In the signal the dependence on
$\lambda$ can be factorized: $R\equiv \lambda \bar{R}$.
An example of the function $\bar{R}$ for several materials
is shown in Fig. 1 as a function of time during
one day and assuming the coordinates of the LNGS laboratory.
The crystallographic inputs in the calculation are given
in the Appendix. In all the plots the main axis of the
lattice have been taken parallel to the South, West and upward directions.
This explains the symmetry shown by all the plots.

The signal is peaked in energy around the maximum of the flux of
Eq.(\ref{eq:flux})
and presents a strong sub--diary
dependence on time, due to the motion of the Sun in the sky.
The time duration of the peaks decreases with growing energies,
from tens of minutes in the lowest part of the axion energy
window, down to a minimum of about one minute in the higher one, and is related to
the energy resolution of each detector.

In order to extract the signal from the background for each energy interval
$E_k<E<E_k+\Delta E$
we introduce, following Ref.\cite{cosmesur}, the quantity:
\begin{equation}
\chi=\sum_{i=1}^{n} \left[ \bar{R}(t_i)-<\bar{R}>\right]\cdot
n_i\equiv \sum_{i=1}^{n} W_{i} \cdot
n_{i}
\label{eq:chi}
\end{equation}
\noindent where the
$n_i$ indicate the number of measured events in the time bin
$t_i,t_i+\Delta t$ and the sum is over
the total period $T$ of data taking. Here and in the following
the brackets indicate time average.

By definition the quantity $\chi$
is expected to be compatible with zero in absence of a signal,
while it weights positively the events
recorded in coincidence with the expected peaks.

The time distribution of $n_i$ is supposed to be given by a Poissonian
with average:
\begin{equation}
<n_i>=\left[ \lambda \bar{R}(t_i)+b \right]\Delta t.
\label{eq:mediaconti}
\end{equation}
Assuming that the background $b$ dominates over the signal
the expected average
and variance of $\chi$ are given by:
\begin{eqnarray}
<\chi>&=&\lambda \cdot A\\
\sigma^2(\chi)&\simeq& b/A
\label{eq:medie}
\end{eqnarray}
\noindent with $A\equiv \sum_i W_i^2\Delta t$.
Each energy bin $E_k,E_k+\Delta E$ with background $b_k$
provides an independent estimate $\lambda_k=\chi_k/A_k$
so that one can get the most probable combined value of $\lambda$:
\begin{eqnarray}
\lambda&=&\sum_k\chi_k/\sum_j A_j \nonumber \\
\sigma(\lambda)&=&\left(\sum_k A_k/b_k\right)^{-\frac{1}{2}}.
\label{eq:lambda}
\end{eqnarray}

The sensitivity of an axion experimental search can be expressed
as the upper bound of $\gagamma$ which such experiment would
provide from the non--appearance of the axion signal, for a given
crystal, background and exposition.
If $\lambda$ is compatible to zero, then at the 95\% C.L.
$\lambda\lsim$ 2$\times$ 1.64$\times\sigma(\lambda)$.

It is easy to see that the ensuing limit on the axion--photon coupling
$\gagamma^{lim}$ scales with the background b (assumed to be flat) and
exposure MT in the following way:
\begin{equation}
\gagamma\leq \gagamma^{lim}\simeq {\rm K}
\left(\frac{\rm b}{{\rm cpd/kg/keV}}\times\frac{\rm kg}
{{\rm M}}\times\frac{\rm years}{\rm T} \right)^{\frac{1}{8}}\times 10^{-9} \;
{\rm GeV}^{-1}
\label{eq:limit}
\end{equation}
\noindent where M is the detector mass and T the total time.
The factor K$\simeq$ 6.5 $\left[\sum_k <(W(E_k)/M)^2>
{\rm keV}/\Delta E_k\right]^{-1/8}$
is a function of
the parameters of the crystal, as well as of the experimental threshold and
resolution.
The application of the statistical analysis described above
results in a background rejection of about two orders of
magnitude.

\section{Discussion and conclusions}

In order to perform a systematic analysis of the axion--detection capability of
crystal detectors, we have applied the technique
described in the previous section to several materials.
The result is summarized in Table \ref{tab:exp}, where
the limit given by the experiment of Ref.\cite{cosmesur}
is compared to those attainable with
running\cite{dama1},
being installed\cite{cuoricino,nai_canfranc} and
planned\protect\cite{cuore1,genius1}
crystal detector experiments.

In Table \ref{tab:exp} a Pb detector is also included,
to give an indication
of the best improvement that one would
expect by selecting heavy materials to take advantage of the proportionality to $Z^2$
of the cross section of Eq.(\ref{eq:cross}).
These results can be easily extended to other elements with lower
atomic numbers than those considered in Table 1
(for instance in Al$_2$O$_3$ or LiF crystals, also in operation or planned dark matter
experiments) although they are expected to yield less stringent limits.

In Fig.2 the result of our analysis is compared to the present
astrophysical and experimental bounds
in the plane $m_a$--$\gagamma$.
The horizontal thick line represents the constraint
$\gagamma\lsim$3$\times$10$^{-10}$ GeV$^{-1}$ taken from Table \ref{tab:exp}.
The
mass intervals $m_a\lsim 10^{-6}$ eV, $10^{-2}$ eV $\lsim m_a \lsim 3$
eV, $m_a\gsim 20$ eV are excluded respectively
by cosmological overclosure\cite{kolbturner}, SN1987A\cite{sn87a}
and oxygen excitation in water Cherenkov detectors\cite{cerenkov},
the edges of all the excluded regions being subject to
many astrophysical and theoretical uncertainties.
The horizontal lines represent the limits from
SOLAX\cite{cosmesur}, helioseismology\cite{helio},
and globular clusters\cite{hb}.
The theoretical predictions for $\gagamma$ in the case of the
DFSZ and the KSFZ axion models are also given.

As shown in the expression of the $\gagamma$ bound of
Eq.(\ref{eq:limit}), the improvement in background and accumulation of statistics
is washed out by the 1/8 power dependence of $\gagamma$ on such
parameters. It is evident, then,
that crystals have
no realistic chances to challenge the globular cluster limit.
A discovery of the axion by this technique would presumably imply
either a systematic effect in the stellar--count observations in globular
clusters or a substantial change in the theoretical models
that describe the
late--stage evolution of low--metallicity stars.

On the other hand,
the sensitivity required for crystal--detectors in order to explore a
range of $\gagamma$,
compatible with the solar limit of Ref.\cite{helio},
appears to be within reach, although only marginally, and
provided that large improvements in background as well as
substancial increase of statistics be guaranteed.

In light of the previous discussion, prospects to detect the axion
with crystal detectors appear to be dim.
The only realistic possibility to exploit this technique seems to rely on the fact
that
collecting a statistics of the order of a few tons$\times$year could be
achievable by adding properly the results of the various
dark matter search experiments already planned for the near future.
As shown in Fig.2 the ensuing limit for crystals would cross the theoretical
models for an axion mass in the range of a few eV, in a region
compatible with the bound given by SN1987A.

\section*{Appendix}
In this section we summarize the crystallographic information necessary to
calculate the axion--photon coherent conversion discussed in section
\ref{sec:theory} for the materials shown in Table \ref{tab:exp}.
We recall that a Bravais lattice
generated by the primitive vectors $\vec{x}_i$ has a reciprocal
lattice generated by the basis $\vec{G}_i=2
\pi/v_a\epsilon_{ijk}\vec{x}_j\times\vec{x}_k$, where
$v_a=\vec{x}_1\cdot (\vec{x}_2\times \vec{x}_3)$ is the volume of
the primitive cell, while $\epsilon_{ijk}$ is the totally anti--commutating
tensor.

Germanium, sodium iodide and lead share the same
type of underlying Bravais lattice, the
faced--centered cubic crystal (fcc) structure generated by the primitive
vectors\cite{cristalografia}:
\begin{eqnarray}
\vec{x}_1&=&\frac{a}{2}(0,1,1) \nonumber \\
\vec{x}_2&=&\frac{a}{2}(1,0,1) \nonumber \\
\vec{x}_3&=&\frac{a}{2}(1,1,0).
\label{eq:generators}
\end{eqnarray}
The corresponding primitive cell sizes $a$ are given in Table
\ref{tab:fcc}, along with the basis vectors for Ge and NaI.

The Bravais lattice of TeO$_2$ is tetragonal.
In the plots of Fig. 2 the longer
side of the cell has been chosen in the vertical
direction. In this case the generators are simply:
\begin{eqnarray}
\vec{x}_1&=&a(1,0,0) \nonumber \\
\vec{x}_2&=&a(0,1,0) \nonumber \\
\vec{x}_3&=&b(0,0,1).
\label{eq:generators_teo2}
\end{eqnarray}
\noindent with a=4.796 ${\rm\AA}$ and b=7.626 ${\rm\AA}$.
Each primitive cell contains a basis of 4
Tellurium atoms and 8 Oxygen atoms. Their positions in the primitive cell
are given in Table \ref{tab:teo2}.

The parameter $r_j$ entering the Yukawa potential of Eq.(\ref{eq:potential})
is of the order of the atomic dimensions.
In our calculation we have assumed $r_j$=1 ${\rm\AA}$ for all the
crystals.
Varying $r_j$ in the range 0.5 ${\rm\AA}\lsim r_j\lsim$ 1 ${\rm\AA}$
affects
the $\gagamma$ limit by less than $\simeq$ 30\%.

\section*{Acknowledgements}
This search has been partially supported by the Spanish Agency of
Science and Technology (CICYT) under the grant AEN98--0627. One of us (S.S.) is
supported by a fellowship of the INFN (Italy).
The authors wish to thank Frank Avignone for useful discussions.

\vfill
\eject

\begin{table}
\centering
\caption{
Axion search sensitivities for running (DAMA\protect\cite{dama1}),
being installed (CUORICINO\protect\cite{cuoricino}, CANFRANC\protect\cite{nai_canfranc})
and planned (CUORE\protect\cite{cuore1}, GENIUS\protect\cite{genius1}) experiments
are compared to the result of SOLAX\protect\cite{cosmesur}. To
estimate the best improvement that would be possible by selecting high--Z materials, a
calculation is also shown for a Pb lattice, assuming  the same conditions of background,
threshold and resolution of the GENIUS experiment.
The coefficient K has been introduced in Eq.(\ref{eq:limit}).
\label{tab:exp}
}
\vspace{0.4cm}
\begin{center}
\begin{tabular}{|c|c|c|c|c|c|c|} \hline
  & {\bf K} & {\bf M} & {\bf b } & {\bf $E_{th}$}&
 {\bf FWHM } & {\bf $\gagamma^{lim}(2\; {\rm years})$} \\
 &         &    {\bf (kg)} & {\bf (cpd/kg/keV)} & {\bf (keV)} & {\bf (keV)} &
 {\bf (GeV$^{-1}$)} \\ \hline
{\bf Ge}\cite{cosmesur} & 2.5 & 1 & 3 & 4 & 1 & 2.7$\times
10^{-9}$\\ \hline
{\bf Ge}\cite{genius1} & 2.5 & 1000 & 1$\times 10^{-4}$ & 4 & 1 & 3$\times
10^{-10}$\\ \hline
{\bf TeO$_2$}\cite{cuoricino}& 3 & 42 & 0.1 & 5 & 2 &
1.3$\times 10^{-9}$ \\ \hline
{\bf TeO$_2$}\cite{cuore1}& 2.8 & 765 & $1\times 10^{-2}$ & 3 & 2 &
6.3$\times 10^{-10}$ \\ \hline
{\bf NaI}\cite{dama1} & 2.7 & 87 & 1 & 2 & 2 & 1.4$\times 10^{-9}$ \\ \hline
{\bf NaI}\cite{nai_canfranc} & 2.8 & 107 & 2 & 4 & 2 & 1.6$\times 10^{-9}$ \\ \hline
{\bf Pb}                & 2.1 & 1000 & 1$\times 10^{-4}$ & 4 & 1 & 2.5$\times 10^{-10}$ \\
\hline
\end{tabular}
\end{center}
\end{table}

\begin{table}
\centering
\caption{
Crystallographic information for Ge, NaI, and Pb:
$a$ is the primitive cell size and the $\vec{a}_i^j$ are the
basis vectors entering in the evaluation of the structure functions
of Eq.(\ref{eq:struct}).
\label{tab:fcc}
}
\vspace{0.4cm}
\begin{center}
\begin{tabular}{|c|c|c|} \hline
 {\bf Species} & {\bf a (${\rm\AA}$)}  & Basis $\vec{a}_i^j$ \\ \hline
 Ge & 5.66 & $\vec{a}_1^1$=(0,0,0)\\
   &  & $\vec{a}_2^1$=$\frac{a}{4}$(1,1,1)\\ \hline
  NaI  & 6.47 & $\vec{a}_1^1$=(0,0,0) (I)\\
     &  &  $\vec{a}_1^2$=$\frac{a}{2}$(1,1,1) (Na)\\ \hline
   Pb   & 4.95 & No basis\\ \hline
\end{tabular}
\end{center}
\end{table}

\begin{table}
\centering
\caption{
Basis vectors for TeO$_2$ entering in the evaluation of the
structure function of Eq.(\ref{eq:struct}). The longer side of the tetragonal
primitive cell has been chosen in the vertical direction.
All numbers are in ${\rm\AA}$.
\label{tab:teo2}
}
\vspace{0.4cm}
\begin{center}
\begin{tabular}{|ll|l|} \hline
 {\bf Oxygen} &  & {\bf Tellurium} \\ \hline
$\vec{a}_1^1$=( 0.998 , 0.916 , 1.480) &
$\vec{a}_2^1$=( 3.797 , 3.879 , 5.293) &
$\vec{a}_1^2$=( 0.093 , 0.093 , 0.000)
\\
$\vec{a}_3^1$=( 1.481 , 3.396 , 3.386) &
$\vec{a}_4^1$=( 3.314 , 1.399 , 7.199) &
$\vec{a}_2^2$=( 4.702 , 4.702 , 3.813)
\\
$\vec{a}_5^1$=( 1.399 , 3.314 , 0.426) &
$\vec{a}_6^1$=( 3.396 , 1.481 , 4.239) &
$\vec{a}_3^2$=( 2.304 , 2.491 , 1.906)
\\
$\vec{a}_7^1$=( 0.916 , 0.998 , 6.145) &
$\vec{a}_8^1$=( 3.879 , 3.797 , 2.332) &
$\vec{a}_4^2$=( 2.491 , 2.304 , 5.719)
\\
\hline
\end{tabular}
\end{center}
\end{table}

\eject
\section*{Figure Captions}

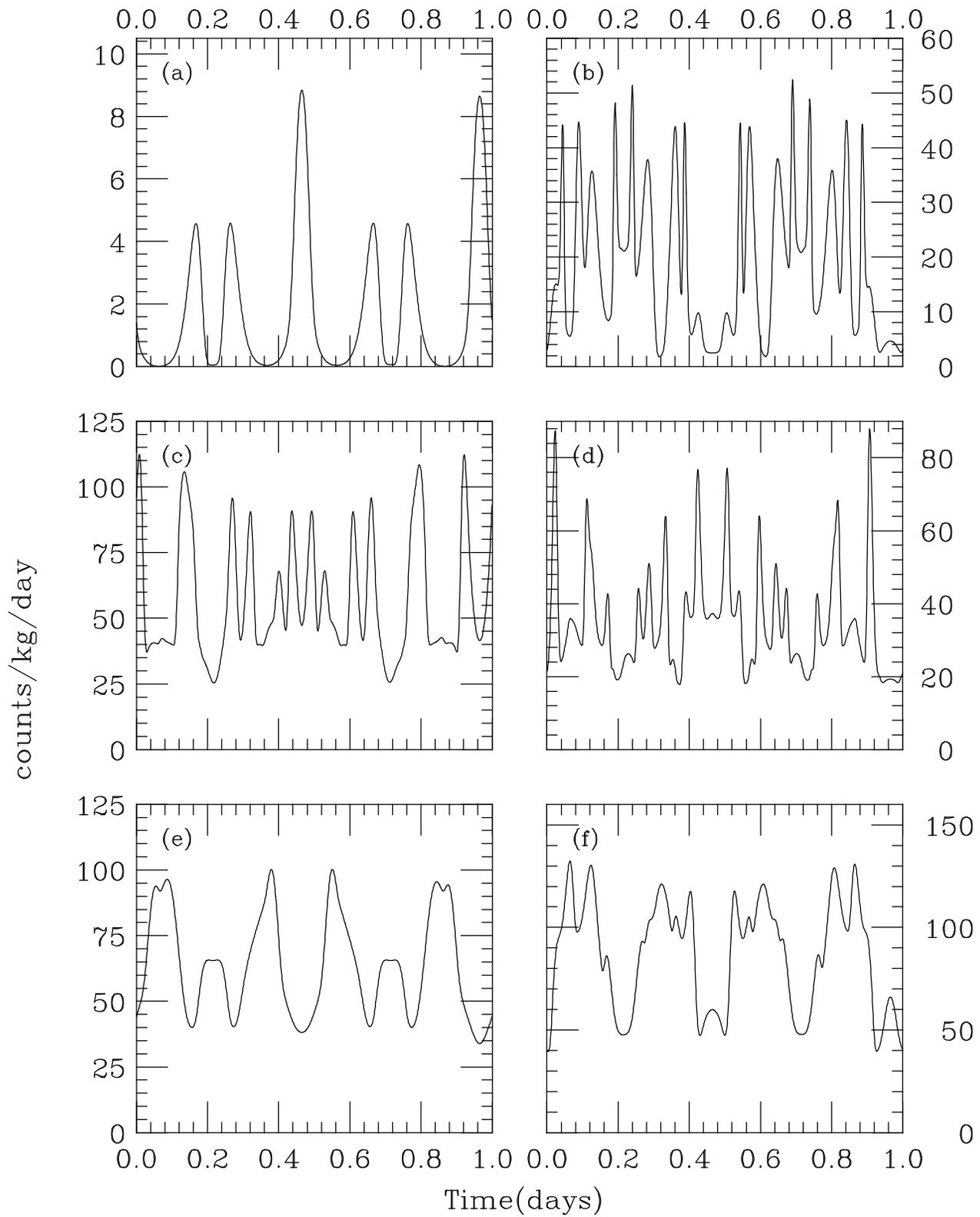
{\bf Figure 1} --
Expected axion signals for Primakoff conversion in various
crystals as a function of time for $\lambda=1$.
In the calculation the representative
day of 1 april 1998 and the coordinates of the LNGS laboratory
have been assumed. From top--left to bottom--right:
a) Ge, 2 keV$\leq E_{ee}\leq$2.5 keV; b) Ge, 4 keV$\leq E_{ee}\leq$4.5 keV;
c) TeO$_2$, 5 keV$\leq E_{ee}\leq$7 keV; d) TeO$_2$, 7 keV$\leq E_{ee}\leq $9 keV;
e) NaI, 2 keV$\leq E_{ee}\leq $4 keV; f) NaI, 4 keV$\leq E_{ee}\leq$6 keV.

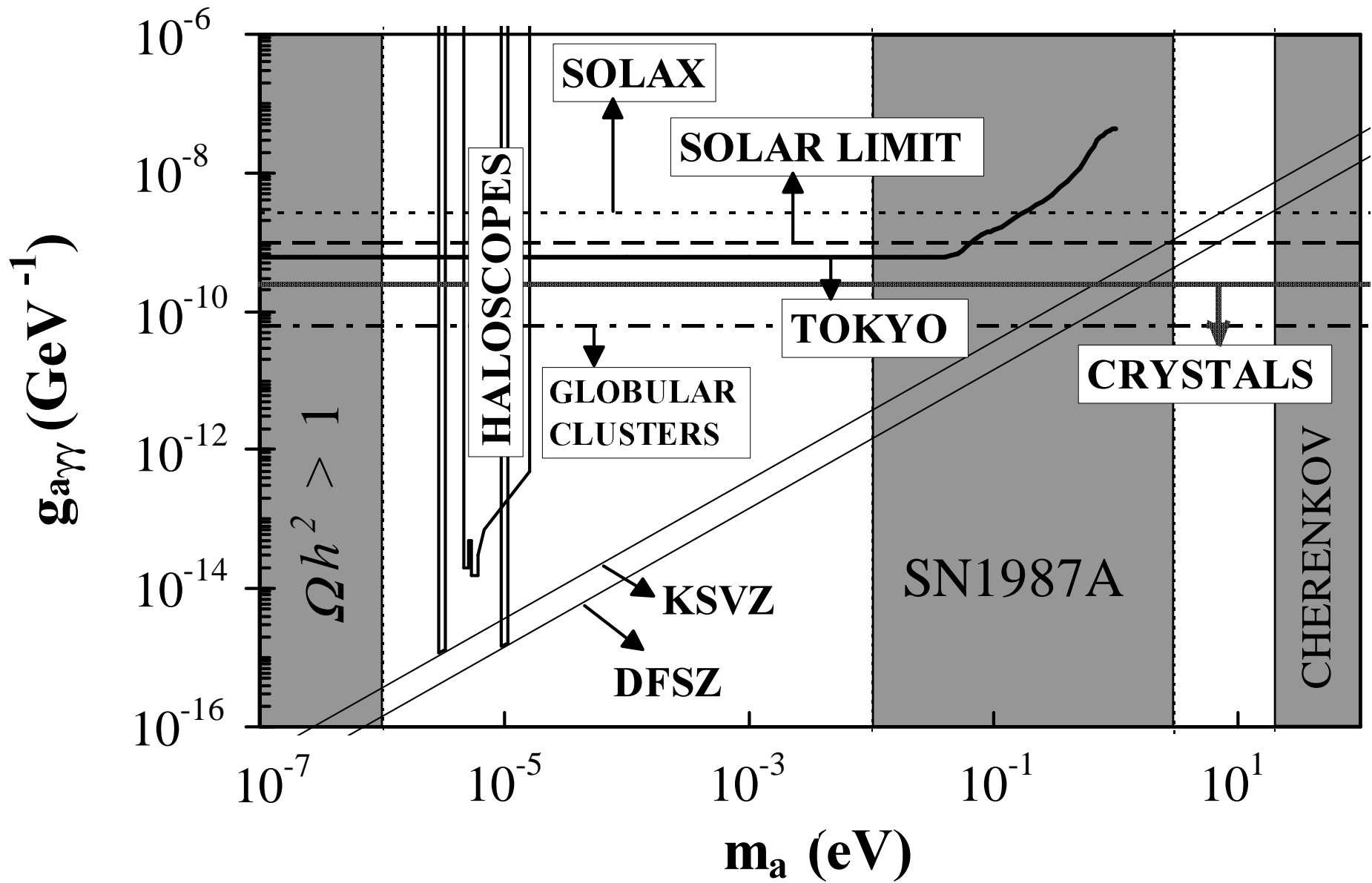
{\bf Figure 2} --
The solar axion limit attainable with crystal detectors
(horizontal thick line) is compared to the present
astrophysical and experimental bounds and to the DFSZ and KSVZ axion
theoretical predictions.
The shaded mass intervals $m_a\lsim 10^{-6}$ eV, $10^{-2}$ eV $\lsim m_a \lsim 3$
eV, $m_a\gsim 20$ eV are excluded respectively by
cosmological overclosure\cite{kolbturner}, SN1987A
\cite{sn87a}
and oxygen excitation in water Cherenkov detectors\cite{cerenkov}.
The horizontal lines represent the upper limits for $\gagamma$
from helioseismology\cite{helio} (dashes), SOLAX\cite{cosmesur} (dots) and
globular clusters\cite{hb}
(dot--dashes). The regions excluded by haloscope searches\cite{cavities1} and the
upper limit on $\gagamma$ from the Tokio Helioscope experiment\cite{tokio1}
are also shown.

\end{document}